\documentstyle[tighten,floats,12pt,prd,aps,eqsecnum]{revtex}

\input epsf
\begin{document}
\date{\today}
\title{Radion effective potential in the Brane-World}

\author{Jaume Garriga$^1$, Oriol Pujol\`as$^1$ and Takahiro Tanaka$^2$
\\ \hspace*{1cm}}

\address{$^1$~IFAE, 
Departament de F\'\i sica, Universitat Aut\`onoma de Barcelona,\\
08193 Bellaterra $($Barcelona$)$, Spain}

\address{$^{2}$~Yukawa Institute for Theoretical Physics, 
Kyoto University, Kyoto 606-8502, Japan
}
\maketitle

\thispagestyle{empty}
\vspace{1cm}
\centerline{\bf Abstract}
\vspace{2mm}
\begin{abstract}
We show that in brane-world scenarios with warped extra dimensions, 
the Casimir force due to bulk matter fields may be sufficient to stabilize
the radion field $\phi$. In particular, we calculate one loop effective 
potential for $\phi$ induced by bulk gravitons and other possible 
massless bulk fields in the Randall-Sundrum background.
This potential has a local extremum, which can be a maximum or a minimum 
depending on the detailed bulk matter content. If the 
parameters of the background are chosen so that the hierarchy 
problem is solved geometrically, then the radion mass induced by Casimir
corrections is hierarchycally smaller than the $TeV$. Hence, in this important
case, we must invoke an alternative mechanism (classical or nonperturbative) 
which gives the radion a sizable mass, to make it compatible with 
observations.
\end{abstract}

\hspace*{1cm}
$~~~~~~~~~~~~~~~~~~~~~~~~~~~~~~~~~~~~~~~~~~~~~~~~~~~~$ YITP-00-20
$~~~~$UAB-FT 486 
\vspace{1cm}
\newpage

\section{Introduction}

It has been suggested that
theories with extra dimensions may provide a solution to the hierarchy
problem \cite{gia,RS1}. The idea is to introduce a $d$-dimensional 
internal space of large physical volume 
${\cal V}$, so that the the effective lower dimensional Planck mass 
$m_{pl}\sim {\cal V}^{1/2} M^{(d+2)/2}$ is much larger than $M \sim TeV$-
the true fundamental scale of the theory. In the original
scenarios, only gravity was allowed to propagate in the higher
dimensional bulk, whereas all other matter fields were confined to live 
on a lower dimensional brane. 
Randall and Sundrum \cite{RS1} (RS) introduced a particularly attractive
model where the gravitational field created by the branes is taken into 
account. Their background solution consists of two parallel flat branes, one 
with positive tension and another with negative tension 
embedded in a a five-dimensional Anti-de Sitter (AdS) bulk. In this model,
the hierarchy problem is solved if the distance between branes 
is about $37$ times the AdS radius and 
we live on the negative tension brane. More recently, 
scenarios where additional fields propagate in the bulk have 
been considered \cite{alex1,alex2,bagger}.

In principle, the distance between 
branes is a massless degree of freedom, the radion field $\phi$.
However, in order to make the theory  compatible with observations 
this radion must be stabilized \cite{gw1,gw2,gt,cgr,tm}. Clearly, 
all fields which propagate in the bulk will give Casimir-type contributions
to the vacuum energy, and it seems natural to investigate whether these
could provide the stabilizing force which is needed.
In this paper, we shall calculate the radion effective potential 
$V_{\hbox{\footnotesize\it \hspace{-6pt} eff\,}}(\phi)$ due to KK gravitons and other
massless bulk fields. As we shall see, this
effective potential has a rather non-trivial behaviour, 
which generically develops a local 
extremum. Depending on the detailed matter content, the 
extremum could be a maximum or a minimum, where the radion could sit. 
For the purposes  of illustration, here we shall concentrate 
on the background geometry discussed by Randall and Sundrum, although 
our methods are also applicable to other geometries, such as the one 
introduced by  Ovrut et al. in the 
context of eleven dimensional supergravity with one large extra dimension
\cite{ovrut}.

The paper is organized as follows. In Section II, we introduce the model
and conventions. In Section III we calculate the effective  
potential due to conformally invariant bulk fields, and discuss the
stabilization mechanism. Conformally invariant fields are convenient 
because of their simplicity, which allows a quick derivation of their
contribution. Moreover, the backreaction of the Casimir 
energy on the geometry can be taken into consideration in this case. 
Section IV is devoted to the graviton contribution. 
For gravitons, conformal invariance is lost, and the evaluation of the 
effective potential is not straightforward. A method will be developed
which relates the curved space effective potential to a suitable effective
potential in flat space. The later takes the form of a sum over the 
contributions of an infinite tower of massive KK fields, whose mass  spectrum
$m_n(\phi)$ is a function of the brane separation (the discrete index 
$n$ labels the infinite KK tower). Our conclusions are summarized in Section V.

Related calculations of the Casimir interaction amongst branes have been 
presented in an interesting paper by Fabinger and Horava \cite{FH}. In 
the concluding section we shall comment on the differences between their 
results and ours.

\section{The Randall-Sundrum model and the radion field}

To be definite, we shall focus attention on the 
brane-world model introduced by Randall and Sundrum \cite{RS1}.
In this model the metric in the bulk is anti-de Sitter space
(AdS), whose (Euclidean) line element is given by 
\begin{equation}
 ds^2=a^2(z)\eta_{ab}dx^{a}dx^{b}=
     a^2(z)\left[dz^2 +d{\bf x}^2\right] 
   =dy^2+a^2(z)d{\bf x}^2.
\label{rsmetric}
\end{equation}
Here $a(z)=\ell/z$, where $\ell$ is the AdS radius.
The branes are 
placed at arbitrary locations which we shall denote by $z_+$ and
$z_-$, where the positive and negative signs refer to the positive and
negative tension branes respectively ($z_+ < z_-$).
The ``canonically normalized'' radion modulus
$\phi$ - whose kinetic term contribution to the
dimensionally reduced action on the positive tension brane is given by 
\begin{equation}
 {1\over 2}\int d^4 x \sqrt{g_+}\, g^{\mu\nu}_+\partial_{\mu}\phi 
    \,\partial_{\nu}\phi, \label{kin}
\end{equation}
-is related to the proper 
distance $d= \Delta y$ between both branes in the following way \cite{gw1}
$$
\phi=(3M^3\ell/4\pi)^{1/2} e^{- d/\ell}.
$$
Here, $M \sim TeV$ is 
the fundamental five-dimensional Planck mass. It is usually assumed
that $\ell \sim M^{-1}$ . Let us introduce the dimensionless radion
$$
\lambda \equiv \left({4\pi \over 3M^3\ell}\right)^{1/2} {\phi} = 
{z_+ \over z_-}   = e^{-d/\ell},
$$
which will also be refered to as {\em the hierarchy}. 
The effective four-dimensional Planck mass $m_{pl}$ 
from the point of view of the negative tension brane is 
given by $m_{pl}^2 = M^3 \ell 
(\lambda^{-2} - 1)$. With $d\sim 37 \ell$, 
$\lambda$ is the small number responsible for the 
discrepancy between $m_{pl}$ and $M$.

At the classical level, the radion is massless. However, as we shall see, 
bulk fields give rise to a Casimir energy which depends on the interbrane
separation. This induces an effective potential $V_{\hbox{\footnotesize\it \hspace{-6pt} eff\,}}(\phi)$ which by
convention we take to be the energy density per unit physical volume on the
positive tension brane, as a function of $\phi$.
This potential must be added 
to the kinetic term (\ref{kin}) in order to obtain the effective action for 
the radion:
\begin{equation}
 S_{\hbox{\footnotesize\it \hspace{-6pt} eff\,}}[\phi]
 =\int d^4x\, a_+^4 \left[{1\over 2}g_+^{\mu\nu}\partial_{\mu}\phi\, 
         \partial_{\nu}\phi +
          V_{\hbox{\footnotesize\it \hspace{-6pt} eff\,}}(\lambda(\phi))
        \right].
\label{effect}
\end{equation}
In the following Sections,  we calculate the contributions to 
$V_{\hbox{\footnotesize\it \hspace{-6pt} eff\,}}$ from
conformally invariant bulk fields and from bulk gravitons.

\section{Radion stabilization by massless bulk fields}

Let us start by considering the contribution to 
$V_{\hbox{\footnotesize\it \hspace{-6pt} eff\,}}(\phi)$  
from conformally invariant massless bulk fields. Technically, this is 
much simpler than finding the contribution from bulk gravitons which will be
discussed in the next section. Moreover 
the problem of backreaction of the Casimir energy onto the background can 
be taken into consideration in this case. 

Consider, for instance, a conformally coupled scalar $\chi$. 
This obeys the equation of motion
\begin{equation}
-\Box_g \chi + {D-2 \over 4 (D-1)}\ R\ \chi =0. 
\label{confin}
\end{equation}
where $\Box_g$ is the d'Alembertian operator in the metric
(\ref{rsmetric}) and $R$ is the Ricci scalar. 
Here, we consider the case 
of arbitrary odd spacetime dimension $D$, with branes of co-dimension one.
By changing the variable 
$\chi \to \hat\chi = a^{(D-2)/2} \chi$, the equation of motion for $\chi$
becomes
\begin{equation}
\bar\Box \hat\chi =0.
\label{fse}
\end{equation}
Here $\bar\Box$ is the {\em flat space} d'Alembertian. It is customary to
impose $Z_2$ symmetry on the bulk fields, which results in Neumann boundary 
conditions
$$
\partial_{z}\hat\chi = 0,
$$
at $z_+$ and $z_-$. 
The eigenvalues of the d'Alembertian subject to these conditions are
given by 
\begin{equation}
\label{flateigenvalues}
\lambda^2_{n,k}=\left({n \pi \over L}\right)^2+k^2,
\end{equation}
where $n$ is a positive integer, $L=z_{-}-z_+$ is the coordinate 
distance between 
both branes and $k$ is the coordinate momentum parallel to the branes.

Similarly, we could consider the case of massless fermions in the RS 
background. The Dirac equation is conformally invariant
\cite{bida}, and the conformally rescaled components of the 
fermion obey the flat space equation (\ref{fse}) with Neumann boundary 
conditions. Thus, the spectrum (\ref{flateigenvalues}) is
also valid for massless fermions.

Let us now consider the Casimir energy density in the conformally 
related flat space problem. We shall first look at the effective potential 
per unit area on the brane, ${\cal A}$. For bosons, this 
is given 
\begin{equation}
V^b_0 = {1\over 2 {\cal A}} {\rm Tr}\ {\rm\ln} (-\bar\Box/\mu^2).
\end{equation}
Here $\mu$ is an arbitrary renormalization scale.
Using zeta function regularization [See e.g. \cite{ramond}, or
the discussion in Section IV for details], it is straightforward to 
show that \begin{equation}
V^b_0 (L)= {(-1)^{\eta-1} \over (4\pi)^{\eta} \eta!}
\left({\pi\over L}\right)^{D-1} \zeta'_R(1-D).
\label{vboson}
\end{equation}
Here $\eta=(D-1)/2$, and $\zeta_R$ is the standard Riemann's zeta function.
The contribution of a massless fermion is given by the same expression 
but with opposite sign:
\begin{equation}
V^{f}_0(L) = - V_0^b(L).
\label{vfermion}
\end{equation}
The expectation value of the energy momentum tensor
is traceless in flat space for conformally invariant 
fields 
%\footnote{Incidentally, 
%this is not true for gravitons, as was assumed in \cite{FH}}. 
Moreover, because of the symmetries of
our background, it must have the form \cite{bida}
$$
\langle T^z_{\ z}\rangle_{flat}= (D-1) \rho_0(z),\quad 
\langle T^i_{\ j}\rangle_{flat}= -{\rho_0(z)}\ \delta^i_{\ j}.
$$
By the conservation of energy-momentum, $\rho_0$ must
be a constant, given by
$$
\rho_0^{b,f} = {V_0^{b,f} \over 2 L} = \mp {A \over 2 L^D},
$$
where the minus and plus signs refer to bosons and fermions respectively and we
have introduced
$$
A\equiv{(-1)^{\eta} \over (4\pi)^{\eta} \eta!}
\pi^{D-1} \zeta'_R(1-D) > 0.
$$

Now, let us consider the curved space case.
Since the bulk is of odd dimension, there is no anomaly 
\cite{bida} and the
energy momentum tensor is traceless in the
curved case too. This tensor is related to the flat 
space one by (see e.g. \cite{bida})
$$
\langle T^{\mu}_{\ \nu}\rangle_g = a^{-D} \langle T^{\mu}_{\ \nu}
\rangle_{flat}.
$$
Hence, the energy density is given by
\begin{equation}
\rho = a^{-D} \rho_0.
\label{dilute}
\end{equation}
The effective potential per unit physical volume on the positive tension
brane is thus given by
\begin{equation}
V_{\hbox{\footnotesize\it \hspace{-6pt} eff\,}}(\lambda) =
2\ a_+^{1-D} \int a^D(z) \rho\, dz = 
\mp \ell^{1-D}{A \lambda^{D-1} \over (1-\lambda)^{D-1}}.
\label{ve1}
\end{equation}
Note that the background solution $a(z)=\ell/z$ has only been used in 
the very last step. 

The previous expression for the effective potential takes into account the
casimir energy of the bulk, but it is not complete because in general 
the effective potential receives additional contributions from both branes. 
As we shall discuss in more 
detail in Section IV, we can always add to 
$V_{\hbox{\footnotesize\it \hspace{-6pt} eff\,}}$ terms which 
correspond to
finite renormalization of the tension on both branes. These are 
proportional to $\lambda^0$ and $\lambda^{D-1}$.
The coefficients in front of these two powers of $\lambda$ 
cannot be determined from our calculation and can only be fixed by 
imposing suitable renormalization conditions which relate them to observables.
Adding those terms and particularizing to the case of $D=5$, we have
\begin{equation}
V_{\hbox{\footnotesize\it \hspace{-6pt} eff\,}}(\lambda) = \mp \ell^{-4}\left[{A\lambda^4 \over (1-\lambda)^4} +
\alpha+\beta\lambda^4\right],
\label{confveff}
\end{equation}
where $A\approx 2.46 \cdot 10^{-3}$.
The values $\alpha$ and $\beta$ can be obtained from the observed value of the 
``hierarchy'', $\lambda_{obs}$, and the observed value of the effective 
four-dimensional cosmological constant, which we take to be zero.
Thus, we take as our renormalization conditions
\begin{equation}
V_{\hbox{\footnotesize\it \hspace{-6pt} eff\,}}(\lambda_{obs})
  ={dV_{\hbox{\footnotesize\it \hspace{-6pt} eff\,}}\over d\lambda}(\lambda_{obs})=0. 
\label{renc}
\end{equation}
If there are other bulk fields, such as the graviton, which give 
additional classical or quantum mechanical contributions to the radion 
potential, then those should be included in 
$V_{\hbox{\footnotesize\it \hspace{-6pt} eff\,}}$.
From the renormalization conditions (\ref{renc}) the unknown coefficients
$\alpha$ and $\beta$ can be found, and then the mass of
the radion is calculable. In Fig.~1 we plot (\ref{confveff}) for a 
fermionic field and a chosen value of $\lambda_{obs}$.

\begin{figure}[t]
\centering
\hspace*{-4mm}
%\leavevmode
\epsfysize=10 cm \epsfbox{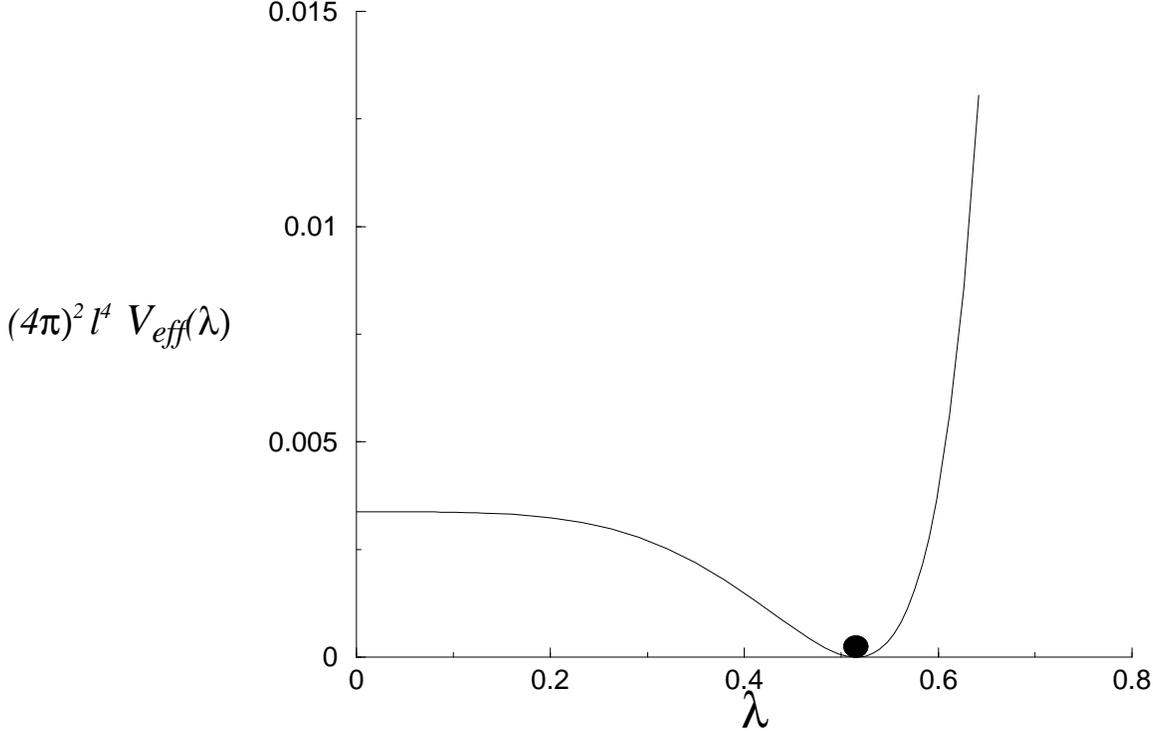}\\[3mm]
%\vspace*{6cm}
\label{fig5}
\caption[fig1]{Contribution to the radion 
effective  potential from a massless bulk fermion. This is plotted 
as a function of the dimensionless radion  $\lambda=e^{-d/\ell}$,
where $d$ is the physical interbrane distance. The 
renormalization conditions (\ref{renc}) have been imposed in order to
determine the coefficients $\alpha$ and $\beta$ which appear in 
(\ref{confveff}).}
\end{figure}

From (\ref{renc}), we have
\begin{equation}
\beta = - A (1-\lambda_{obs})^{-5},\quad \alpha= -\beta 
\lambda_{obs}^5.
\label{consts}
\end{equation}
These values correspond 
to changes $\delta \sigma_{\pm}$ on the positive and negative
brane tensions which are related by the equation
\begin{equation}
\delta\sigma_+ =  -\lambda^5_{obs}\ \delta\sigma_-.
\label{reltensions}
\end{equation}
As we shall see below, Eq. (\ref{reltensions}) is just 
what is needed in order to have a static solution according
to the five dimensional equations of motion, once the casimir 
energy is included.

We can now calculate the 
mass of the radion field $m_\phi^{(-)}$ from the point of view of the negative 
tension brane. For $\lambda_{obs}\ll 1$ we have:
\begin{equation}
m^{2\ (-)}_\phi = 
\lambda_{obs}^{-2}\ m^{2\ (+)}_\phi = \lambda_{obs}^{-2}\ 
   {d^2 V_{\hbox{\footnotesize\it \hspace{-6pt} eff\,}}\over
  d\phi^2}\approx \mp
\lambda_{obs} \left({5\pi^3 \zeta'_R(-4)\over 6 M^3 l^5}\right).
\label{massconf}
\end{equation}
The contribution to the radion mass squared is negative for bosons and 
positive for fermions. Thus, depending on the matter content of the 
bulk, it is clear that the radion may be stabilized due to this effect.

Note, however, that if the ``observed'' interbrane separation is large,
then the induced mass is small. So if we try to solve the
hierarchy problem geometrically with a large internal volume, 
then $\lambda_{obs}$ is of order $TeV/m_{pl}$ and the mass 
(\ref{massconf}) is much smaller than the $TeV$ scale. Such a light 
radion would seem to be in conflict with observations.  
In this case we must accept the existence of 
another stabilization mechanism (perhaps classical or 
nonperturbative) contributing a large mass to the radion. 
Of course, another possibility is to have $\lambda_{obs}$ of order one, 
with $M$ and $\ell$ of order $m_{pl}$, in which case the radion mass
(\ref{massconf}) would be very large, but then we must look for a 
different solution to the hierarchy problem. 

Due to conformal invariance, it is straightforward to take into account the 
backreaction of the Casimir energy on the geometry. First of all, we note 
that the metric (\ref{rsmetric}) is analogous to a
Friedmann-Robertson-Walker metric, where the nontrivial direction is space-like
instead of timelike. The dependence of $a$ on 
the transverse direction can be found from the Friedmann equation
\begin{equation}
\left({a'\over a}\right)^2 = {16\pi G_5 \over 3} \rho - {\Lambda \over 6}.
\label{friedmann}
\end{equation}
Here a prime indicates derivative with
respect to the proper coordinate $y$ [see eq. (\ref{rsmetric})], 
and $\Lambda<0$ is
the background cosmological constant.
Combined with (\ref{dilute}), which relates the energy density $\rho$
to the scale factor $a$, Eq. (\ref{friedmann}) becomes a first order 
ordinary differential equation for $a$. We should also take into account the
matching conditions at the boundaries
\begin{equation}
\left({a'\over a}\right)_{\pm}={\mp 8\pi G_5 \over 6} \sigma_{\pm}. 
\label{matching}
\end{equation}
A static solution of Eqs. (\ref{friedmann}) and (\ref{matching}) can
be found by a suitable adjustment of the brane tensions. Indeed, since the 
branes are flat, the value of the scale factor on the positive tension
brane is conventional and we may take $a_+=1$. Now, the tension $\sigma_+$
can be chosen quite arbitrarily. Once this is done, 
Eq. (\ref{matching}) determines the derivative $a'_+$, and
Eq. (\ref{friedmann}) determines the value of $\rho_0$.
In turn, $\rho_0$ determines the 
co-moving interbrane distance $L$, and hence the location of the second brane.
Finally, integrating (\ref{friedmann}) up to the second brane, 
the tension $\sigma_-$  must be adjusted so that the matching 
condition (\ref{matching}) is satisfied. 
Thus, as with other stabilization scenarios, a single fine-tuning is 
needed in order to obtain a vanishing four-dimensional cosmological constant.

This is in fact the dynamics underlying our 
choice of renormalization conditions (\ref{renc}) which we used in order
to determine $\alpha$ and $\beta$. 
Indeed, let us write
$\sigma_+=\sigma_0 + \delta\sigma_+$ and $\sigma_- =-\sigma_0 +\delta\sigma_-$,
where $\sigma_0=(3 / 4\pi \ell G_5)$ is the absolute value of the 
tension of the branes in the zeroth order background solution. Elliminating
$a'/a$ from (\ref{matching}) and (\ref{friedmann}), we easily recover 
the relation (\ref{reltensions}), which had previously been obtained by 
extremizing the effective potential and imposing zero effective 
four-dimensional cosmological constant (here, $\delta\sigma_{\pm}$
is treated as a small parameter, so that extremization of the effective
action coincides with extremization of the effective potential on the 
background solution.) In that picture, the necessity of a single fine 
tuning is seen as follows. The tension on one of the walls can be chosen 
quite arbitrarily. For 
instance, we may freely pick a value for $\beta$, which renormalizes 
the tension of the brane located at $z_-$. Once this is given, 
the value of the interbrane distance $\lambda_{obs}$ is fixed by the first of
Eqs. (\ref{consts}). Then, the value of $\alpha$, which renormalizes 
the tension of the brane at $z_+$, must be fine-tuned to satisfy the second of
Eqs. (\ref{renc}).

Eqs. (\ref{friedmann}) and (\ref{matching}) can of course be solved 
nonperturbatively. We may consider, for instance, the situation where 
there is no background cosmological constant ($\Lambda=0$). In this case 
we easily obtain
\begin{equation}
a^3(z)={6\pi G A \over (z_- -z_+)^5}(C-z)^2
\end{equation}
where the brane tensions are given by 
$2\pi G \sigma_{\pm}=\pm (C-z_{\pm})^{-1}$ and $C$ is a constant.
This is a self-consistent solution where the warp in the extra dimension
is entirely due to the Casimir energy.

\section{Graviton contribution to 
the radion effective potential} 
%$V_{\hbox{\footnotesize\it \hspace{-6pt} eff\,}}(\phi)$}

Since the graviton is not conformally invariant, the techniques needed 
in order to find its contribution to the effective potential will have 
to be somewhat more sophisticated. However, because of the relevance of 
the graviton, it seems worthwile to pursue 
this task and in the following subsections we develop a convenient
method to address this calculation. This consists of two steps. 
In Subsection IV.A we relate the 
curved space effective potential to a suitable flat space determinant. 
The calculation of this determinant is then presented in Section IV.B.

Our background has maximally symmetric
foliations orthogonal to the $y$ direction, and just like in the 
case of cosmological Friedmann-Robertson-Walker models,
each graviton polarization contributes as a massless minimally 
coupled scalar field. Although this seems to be a well known 
fact which is often quoted in the literature, it is not 
always clear from the context what is the extent of this equivalence.
The correspondence is straightforward at the classical level. At the 
quantum level, it can also be shown to hold, 
although this is not so straightforward to prove because 
careful gauge-fixing of the gravitational sector must be done.
A detailed justification is lengthy, and will be presented elsewhere
\cite{tamaproof}.

In short, the correspondence is as follows.
Perturbations of the gravitational field in the Randall-Sundrum gauge can be 
expressed as 
\begin{equation}
 h_{ab}
  =g_{ab}-g^{(0)}_{ab}=\sum a^2 \sigma_{ab}^{(i)}\Phi_{(i)}, 
\end{equation}
where $\sigma_{55}=\sigma_{\mu\nu}=0$ and $\sigma_{\mu\nu}$
is a constant transverse and traceless polarization 
tensor (a Fourier decomposition of $\Phi$ is assumed along the 
branes, in order to
define transverse polarization). The summation is taken over all 
polarizations. The quadratic reduced action for one particular polarization 
becomes
\begin{equation}
W=\int d^D x \sqrt{g} \Phi \left(-\Box_{g}\Phi\right) + 
(\mbox{boundary term}),  
\label{start}
\end{equation}
where $\Box_{g}$ is the usual covariant 
scalar Laplacian associated with the five dimensional metric
$$ 
g_{ab}=a^2\eta_{ab},$$
and we omitted the index $(i)$. 

Thus, we shall be interested in the determinant of 
the operator  
\begin{equation}
 P=-\Box_{g},
\label{Pdef}
\end{equation} 
with appropriate boundary conditions at the branes. These are 
determined from Israel's junction conditions plus the requirement of 
$Z_2$ symmetry \cite{RS2}. Equivalently, they can be found from the
equation of motion for $\Phi$ in the RS background. It is
easy to show that in terms of $\Phi$, they reduce to the 
standard Neumann boundary conditions i.e.,  
\begin{equation}
 \partial_z\Phi=0, 
\end{equation}
at $z=z_{\pm}$. 

\subsection{From flat to curved space}

Our next task is to evaluate the determinant of the covariant
d'Alembertian operator $P$ 
defined in (\ref{Pdef}). Formally, this is the product of its eigenvalues. 
Nevertheless, in the present case, a straightforward
calculation involving the eigenvalues of $P$ seems rather
complicated and not particularly illuminating.
For instance, the eigenvalues of $P$ are completely 
unrelated to the spectrum of Kaluza-Klein gravitons, 
and so the contribution from individual physical modes seems hard to 
identify.

A more promising strategy
is to express $(\det P)$ in terms of the determinant of an
operator whose eigenvalues are directly related to the KK masses. 
For this purpose, we introduce a one parameter family of metrics
which interpolates between flat space and our physical space,
$$
\tilde g^{\epsilon}_{ab}=\Omega_{\epsilon}^2 g_{ab}.
$$
For definiteness we shall take
\begin{equation}
 \Omega_{\epsilon}(z)={z\over \epsilon z+(1-\epsilon)}, 
\label{conformali}
\end{equation}
where the parameter $\epsilon$ runs from $0$ to $1$.
For $\epsilon=0$ the fictitious metric $\tilde g_{ab}$ is flat
space, whereas for $\epsilon=1$ 
it coincides with the physical metric . The arguments presented 
in this section are independent of the precise form of the ``path''
(see Fig. 2).
The explicit form (\ref{conformali})
will only be used at the very end of the calculation, and it is chosen for  
convenience so that all metrics in the path are 
AdS spaces with curvature radius $\ell/\epsilon$.

\begin{figure}[t]
\centering
\hspace*{-4mm}
%\leavevmode
\epsfysize=10 cm \epsfbox{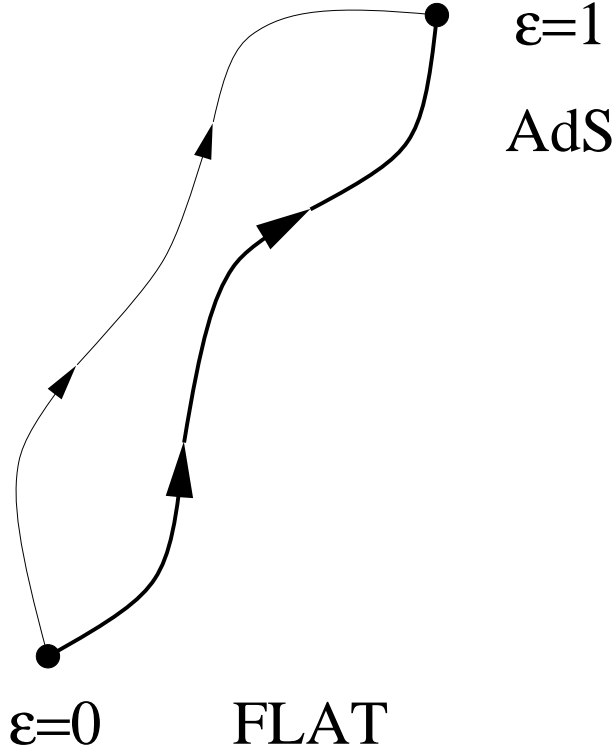}\\[3mm]
%\vspace*{6cm}
\label{fig1}
\caption[fig1]{Paths from flat to AdS space in the space
of conformal factors. The integral in (\ref{conformal}) is 
independent on the path.}
\end{figure}

Next, we introduce the operator $P_{\epsilon}$ defined by
\begin{equation}
\Omega_\epsilon^{\frac{D-2}{2}}{P_\epsilon}
\Omega_\epsilon^{\frac{2-D}{2}}=
\Omega_{\epsilon}^{-2} P
    = -{1\over \Omega_\epsilon^2 a^2}
           \left(z^{D-2}\partial_z z^{2-D}\partial_z 
           +\sum_{i=1}^{D-1}\partial_{i}^2 \right).
\label{allsorts}
\end{equation}
This operator can be written in covariant form
as
$$
P_{\epsilon}= - (\Box_{\tilde g}+ E_{\epsilon}),
$$
where $\Box_{\tilde g}$ is the d'Alembertian in the 
space with metric $\tilde g_{ab}$ and 
$$
{E_{\epsilon}}=\frac{D-2}{4\Omega_\epsilon^2}
   \left[2 \frac{\Box_g \Omega_\epsilon}{\Omega_\epsilon}+(D-4)
 g^{ab}(\partial_a \ln\Omega_\epsilon)(\partial_b \ln\Omega_\epsilon)\right].
$$
For later reference, we note that $P_\epsilon$ acts on the
rescaled field 
\begin{equation}
\tilde \Phi = \Omega_\epsilon^{\frac{2-D}{2}} \Phi,
\label{phitilde}
\end{equation}
which obeys the boundary condition
$
{\cal B}_{\epsilon}\tilde\Phi=
(\tilde n^{a} \partial_{a} + 
S_\epsilon)\tilde\Phi|_{\partial M} =0.
$
Here
\begin{equation}
S_\epsilon= \frac{D-2}{2} \tilde n^{a}
\partial_{a}\ln\Omega_\epsilon,
\label{boundarys}
\end{equation}
and $\tilde n$ is the inward directed normal vector at the boundary.

We are interested in the effective potential per unit co-moving area
${\cal A}_c$. This is given by 
$V^c_{\hbox{\footnotesize\it \hspace{-6pt} eff\,}}= {\cal A}_c^{-1}\ln(\det P)^{1/2}.$
As we shall see, the dependence of $(\det P_\epsilon)$ on $\epsilon$
along the conformal path in Fig. 2 is easily found in terms of 
local quantities. Hence, instead of dealing directly with $(\det P)$,
we shall calculate the flat space determinant $(\det P_0)$, and then
integrate the dependence on $\epsilon$ along the path. The effective 
potential per unit comoving area is given by 
$$
V^c_{\hbox{\footnotesize\it \hspace{-6pt} eff\,}}=V_{\epsilon=1},
$$ 
where
\begin{equation}
 V_{\epsilon}\equiv{1\over 2{\cal A}_c} {\rm Tr}\,{\ln}
         \left({P_{\epsilon}\over \mu^2}\right)
  = -{1\over 2 {\cal A}_c}\lim_{\nu\to 0}\partial_{\nu}
      \zeta_{\epsilon}(\nu).
\label{vepsilon}
\end{equation}
Here, we have introduced
\begin{equation}
 \zeta_{\epsilon}(\nu)= {\rm Tr}\left({P_{\epsilon}\over \mu^2}
              \right)^{-\nu}
  = {2\mu^{2\nu}\over \Gamma(\nu)}\int_0^{\infty}
      {d\xi\over \xi} \xi^{2\nu} 
         {\rm Tr}\left[e^{-\xi^2 P_{\epsilon}}\right]. 
\end{equation}
The symbol ${\rm Tr}$ refers to the usual $L_2$ trace. For any operator
${\cal O}$, the trace can be represented as
$$
{\rm Tr}[{\cal O}]=
\sum_i\int d^D x\ { g}^{1/2}\  \Phi_i ({\cal O}\Phi_i)=
\sum_i\int d^D x\ {\tilde g}^{1/2}\  \tilde\Phi_i ({\cal O}\tilde\Phi_i),
$$
where $\Phi_i$ (or $\tilde \Phi_i$) form an orthonormal basis with respect
to the measure associated with the metric $g$ (or $\tilde g$ respectively).
It is straightforward to 
show that
\begin{equation}
{\rm Tr}\left[f e^{-\xi^2 P_\epsilon}\right]=
{\rm Tr}\left[ f e^{-\xi^2 \Omega_\epsilon^{-2} P}\right].
\label{dostraces}
\end{equation}

Following the Ref.\cite{bg,seeley,gilkey,mks}, we introduce the asymptotic expansion of the trace
for small $\xi$:
\begin{equation}
 {\rm Tr}\left[ f(x) e^{-\xi^2 P_{\epsilon}}\right]
    \sim \sum_{n=0}^{\infty} \xi^{n-D} a_{n/2}(f,P_{\epsilon}),
\label{sdw}
\end{equation}
where $f$ is an arbitrary smooth function.
This is a generalization of the more widely used De Witt-Schwinger expansion 
for the solution of the heat kernel equation \cite{dw}(which 
corresponds to the case $f=1$). 
The dependence of $V_{\epsilon}$ on $\epsilon$ is
related to the coefficient $a_{D/2}$ in the following way
\begin{eqnarray}
 \partial_{\epsilon}\lim_{\nu\to 0}\partial_{\nu}\zeta_{\epsilon}(\nu)
   &=&
      \lim_{\nu\to 0}\partial_{\nu}{2\mu^{2\nu}\over\Gamma(\nu)}
         \int_0^{\infty} d\xi\, \xi^{2\nu} 
      {\rm Tr}\left( -f_{\epsilon} \partial_{\xi}
         e^{-\xi^2 P_{\epsilon}}\right)\cr
   &=& 
      \lim_{\nu\to 0}\partial_{\nu}{4\nu \mu^{2\nu}\over \Gamma(\nu)}
         \int_0^{\infty} d\xi\, \xi^{2\nu-D-1}\left[\xi^{D} 
       {\rm Tr}\left(f_{\epsilon} 
         e^{-\xi^2 P_{\epsilon}}\right)\right]\cr
   &\sim& 
      \lim_{\nu\to 0}\partial_{\nu}{2 \mu^{2\nu}\Gamma^{-1}(\nu)\over 
           (D-2\nu)(D-1-2\nu)\cdots (1-2\nu)}
         \int_0^{\infty} d\xi\, \xi^{2\nu} \partial_\xi^{D+1}
         \sum_{n=0}^{\infty} \xi^{n} a_{n/2}(f_{\epsilon},P_{\epsilon})\cr
   &=& 2 a_{D/2}(f_{\epsilon},P_{\epsilon}), 
\end{eqnarray}
where we have defined 
\begin{equation}
 f_{\epsilon}=\partial_{\epsilon} \ln \Omega_\epsilon.    
\end{equation}
In the first equality, we have used (\ref{dostraces}). 
In the second equality, we performed integration by parts assuming 
$2 \nu > D$. The contribution from $\xi=\infty$ vanishes due 
to the exponential suppression in
${\rm Tr}\left[ f(x) e^{-\xi^2 P_{\epsilon}}\right]$. 
In the third equality, we performed a further integration by parts.
 
The dependence in $\epsilon$ can now be integrated along the path which 
joins flat space with AdS, and the effective potential can be expressed as
\begin{equation}
 V^c_{\hbox{\footnotesize\it \hspace{-6pt} eff\,}}=V_{1}=V_0-{1\over {\cal A}_c}\int_0^1
        d\epsilon\, a_{D/2}(f_\epsilon,P_{\epsilon}). 
\label{conformal}
\end{equation} 
The calculation of $V_0$ is done in the next section. 
The coefficient $a_{D/2}$, for $D\leq 5$, has been studied in
\cite{bg,mks,bgkv,kcd,md,vass} for a large class of ``covariant'' operators
which includes our $P_{\epsilon}$. In general, this can be 
expressed in terms of integrals over the bulk and
boundary of local invariants constructed from the
metric $\tilde g$ and the functions $S_{\epsilon}$ 
and $E_{\epsilon}$ which enter
the boundary condition, given in Eq. (\ref{boundarys}).

For the case of odd dimension $D$, the coefficients $a_{D/2}$ 
contain only contributions from the boundary. For instance, 
for $D=3$ we have \cite{bgkv}
\begin{eqnarray*}
a_{3/2}(f,P_\epsilon) = \frac{1}{1536 \pi}
       {\int_{\partial {\cal M}}} \{&& f(96 E_{\epsilon}+16 \tilde R -8 
   \tilde R^{\mu}{}_{n\mu n} +13 \tilde K^2+ 2 \tilde K_{\mu\nu}
\tilde K^{\mu\nu}\cr
 &&-96 S_\epsilon \tilde K +192S_\epsilon^2)+
f_{ ;n}(-6 \tilde K+96 \tilde S)+24 f_{ ;nn} \}.
\label{sdw}
\end{eqnarray*}
Here $\tilde K_{\mu\nu}$ is the extrinsic curvature and 
$\tilde R^{a}_{bcd}=+\tilde\Gamma^{a}_{bd,c}-...$ is the Riemann tensor. 
As usual,
greek indices run from 0 to 3 and the index $n$ indicates contraction
with the inward unit normal vector $\tilde n$. 
If we integrate this expression along the path given by (\ref{conformal}),
then all intermediate geometries are AdS of radius $\ell/\epsilon$ and 
the invariants are easily found. The result is given by
\begin{equation}
{1\over {\cal A}_c}
\int_0^1 d\epsilon\, a_{3/2}(f_\epsilon,P_\epsilon)=
\frac{1}{4 \pi z_{+}^{2}}\left[\frac{1}{16}(\ln
  z_{+}+\lambda^2 \ln z_{-})-\frac{3}{64}(1+\lambda^2)\right],
\label{int1}
\end{equation}
where we have reintroduced the dimensionless radion $\lambda=z_+/z_-$.
The general expression for $a_{5/2}$, 
relevant to the five dimensional case, has been given by Kirsten 
\cite{klaus}. It is rather lengthy since it involves 
combinations of geometric invariants that contain 4 derivatives of 
the metric, and we shall not reproduce it here. Integrating
this coefficient  over $\epsilon$, we obtain
\begin{equation}
{1 \over {\cal A}_c}\int_0^1 d\epsilon\, a_{5/2}(f_\epsilon,P_\epsilon)
=\frac{1}{4 \pi^2 z_{+}^{4}}\left[
-\frac{27}{128}(\ln  z_{+}+\lambda^4 \ln z_{-})
 +\frac{53}{5120}(1+\lambda^4)\right]. \label{int2}
\end{equation}
It should be noted that under the rescaling 
$z_{\pm}\to \vartheta z_{\pm}$ (with $\vartheta=const.$),
which corresponds to a translation of both branes in 
the $y$ direction, the expression for the effective potential 
$V^c_{\hbox{\footnotesize\it \hspace{-6pt} eff\,}}$ 
must scale like 
\begin{equation}
V^c_{\hbox{\footnotesize\it \hspace{-6pt} eff\,}}\to \vartheta^{1-D} V^c_{\hbox{\footnotesize\it \hspace{-6pt} eff\,}}.
\label{scaling}
\end{equation} 
This shift is a mere coordinate transformation which
changes the physical volume corresponding 
to a given co-moving volume.
We shall show in the next section that the quantity $V_0$ does not scale 
in this way, and that the integrals (\ref{int1}) and
(\ref{int2}) have precisely the form which restores 
the scaling (\ref{scaling}). For that reason,
the explicit evaluation of these integrals is not strictly necessary,
although of course it provides a valuable consistency check.
As we shall see, the integrals can be guessed from 
the scaling property (\ref{scaling}) up to terms which can be reabsorbed by 
finite renormalization.

\subsection{Computation of $V_{\hbox{\footnotesize\it \hspace{-6pt} eff\,}}(\phi)$}

From (\ref{allsorts}),
the eigenvalue problem for $\epsilon=0$ takes the form:
\begin{equation}
\ell^{-2}\left[z^{D-2}\partial_z z^{2-D}\partial_z -k^2\right]
  \Phi_i=-\lambda_i\Phi_i,
\end{equation}
with boundary condition $\partial_z \Phi_i=0$ at $z=z_{\pm}$. 
Here, an expansion in Fourier modes of wavenumber $k$ along the
flat branes has been assumed. In terms of the rescaled field $\tilde \Phi$
defined in (\ref{phitilde}), this equation looks like a Schrodinger equation
with a ``volcano''-type  potential \cite{RS2}. 
Defining $m_i^2=-k^2+\ell^2\lambda_i$, we have 
\begin{equation}
\left[ z^{D-2}\partial_z z^{2-D}\partial_z+m_i^2\right]\Phi_i=0,
\end{equation}
which has a general solution 
$\Phi_i=z^{\eta} \left(A_1 J_{\eta}(m_i z)+
               A_2 Y_{\eta}(m_i z)\right)$. 
The boundary conditions give an equation which determines 
$m_i$ as 
\begin{equation}
 F(\tilde m_i)=J_{\eta-1}(\tilde m_i \lambda)Y_{\eta-1}(\tilde m_i)
        -Y_{\eta-1}(\tilde m_i \lambda)J_{\eta-1}(\tilde m_i)=0, 
\label{mspectrum}
\end{equation}
where $J$ and $Y$ are Bessel functions, $\eta=(D-1)/2$, 
$\tilde m_i= m_i z_-$ and we have reintroduced $\lambda=z_+/z_-$. 

Equation (\ref{mspectrum}) gives the physical spectrum of
masses $m_-=\tilde m_i/\ell$ 
for the KK gravitons from the point of view of the
negative tension brane. This spectrum is a function of the interbrane
distance, which enters Eq. (\ref{mspectrum}) through the hierarchy $\lambda$.
The zeta function for the operator $P_0$, given by
\begin{equation}
 \zeta_{0}(\nu)={\cal A}_c\int {d^{D-1} k\over (2\pi)^{D-1}}
        \sum_{i}\left({k^2+m_i^2\over \ell^2 \mu^2}\right)^{-\nu}, 
\label{yjw}
\end{equation}
resembles the zeta function for a tower of fields of mass $m_i$ in 
flat space. Nevertheless, this is to some extent just a formal analogy:
instead of physical momentum and mass, the
coordinate momenta $k$ and ``coordinate'' mass eigenvalue $m_i$
appear in this expression. \footnote{In fact, it does 
not seem possible to write down an expression analogous to 
(\ref{yjw}) for our $V_{\hbox{\footnotesize\it \hspace{-6pt} eff\,}}$, containing physical masses and momenta
instead of coordinate ones. 
This would be possible if our space-time were 
ultrastatic (see e.g. \cite{bvw}), which just means
that the lapse function is a constant. However, in our case, time flows
at different rates in different places, so for instance the
physical mass depends on which brane we are considering. This seemingly
trivial complication prevents us from doing the naive 
``projection'' of the problem 
onto four dimensional slices, and renormalizing infinities as if we 
were considering an infinite tower of fields living in a flat space.
In fact, such naive procedure would give a qualitatively different result for 
$V_{\hbox{\footnotesize\it \hspace{-6pt} eff\,}}(\phi)$ depending on whether we compute it on the positive 
or on the negative tension brane, which is clearly nonsensical.
This is why the techniques of Section IV.A have been introduced.}

Performing the momentum integrals in (\ref{yjw}) we have
\begin{equation}
\zeta_{0}(\nu) = {\cal A}_c {(\ell\mu)^{2\nu} z_{-}^{2\nu-(D-1)}
             \Gamma(\nu-\eta)
             \over (4\pi)^{\eta}\Gamma(\nu)}\hat \zeta(2\nu-(D-1)),
\label{z0}
\end{equation}
where we defined 
\begin{equation}
   \hat \zeta(s)= \sum_i \tilde m_i^{-s}.
\label{zh}
\end{equation}
Substituting (\ref{z0}) in (\ref{vepsilon}) we have
\begin{equation}
 V_0(\lambda)  =  {(-1)^{\eta-1}\over 
        (4\pi)^\eta \eta! z_-^{D-1}}\left[
       \left\{\ln(\ell\mu z_-)+{1\over 2}
             \sum_{n=1}^{\eta}n^{-1}\right\}\hat\zeta(1-D) 
          +\hat\zeta'(1-D)\right]. 
\label{V0}
\end{equation}
Now the problem reduces to the calculation of the generalized
zeta function (\ref{zh}). This is a discrete sum over the zeros
of the function $F$ defined in (\ref{mspectrum}). This type of 
sums, where $F$ is a combination of Bessel functions, have been
previously studied in \cite{lr1,lr2,elr,peter}. Following
\cite{lr2}, we express $\hat\zeta(s)$ as
\begin{equation}
 \hat\zeta(s) ={1\over 2\pi i}\int_{\cal C} dt\, t^{-s}{F'\over F}
    ={s\over 2\pi i}\int_{\cal C} dt\, t^{-1-s}\ln F.
\label{conver} 
\end{equation}
This follows from the fact that $F$ has only simple zeros and these
are on the positive real axis. The contour
of integration $\cal C$ is given in Fig. 3. 
The expression (\ref{conver}) is convergent for sufficiently large $s$. 

\begin{figure}[t]
\centering
\hspace*{-4mm}
%\leavevmode
\epsfysize=10 cm \epsfbox{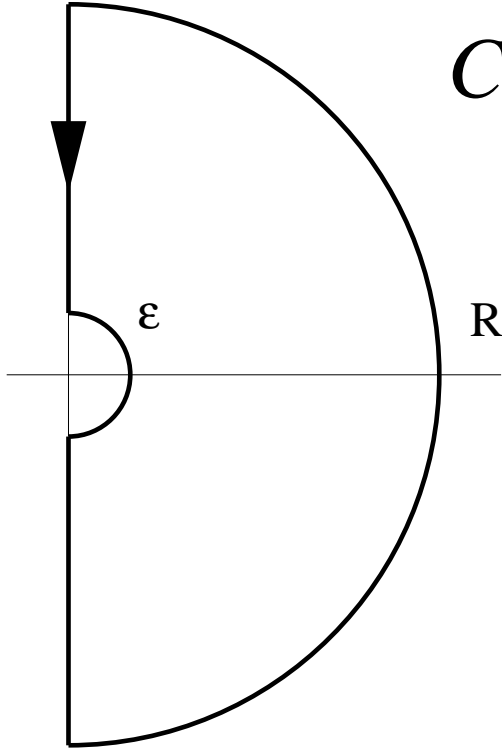}\\[3mm]
%\vspace*{6cm}
\label{fig2}
\caption[fig2]{Contour of integration in Eq. (\ref{conver}).}
\end{figure}

After some manipulations, which are deferred to Appendix A, we find
\begin{eqnarray}
 V_0(\lambda)={1\over (4\pi)^{\eta}(\eta-1)!\, z_+^{D-1}}
\Biggl[&&{\cal I}_K+ \lambda^{D-1} {\cal I}_I+          
          {\beta_{D-1}} (1+\lambda^{D-1})\left(\ln(\ell\mu)+{1\over 2}
                \sum_{n=1}^{\eta-1}n^{-1}\right)
\cr &&
           +\beta_{D-1}\left(\ln z_+ +{\lambda^{D-1}}\ln z_-\right)
         +\lambda^{D-1} {\cal V}(\lambda) \Biggr].
\label{result}
\end{eqnarray}
Here ${\cal I}_I$,${\cal I}_K$ and $\beta_{D-1}$ are constants which are
evaluated in the Appendix, and 
\begin{equation}
{\cal V}(\lambda)=\int_0^{\infty} d\rho\,\rho^{D-2} 
         \ln\left(1-{I_{\eta-1}(\lambda\rho)\over K_{\eta-1}(\lambda\rho)}
                 {K_{\eta-1}(\rho)\over I_{\eta-1}(\rho)}\right),
\label{calvo}
\end{equation}
where $I$ and $K$ are the modified Bessel functions. The function
${\cal V}(\lambda)$ is plotted in Fig. 4.

\begin{figure}[t]
\centering
\hspace*{-4mm}
%\leavevmode
\epsfysize=10 cm \epsfbox{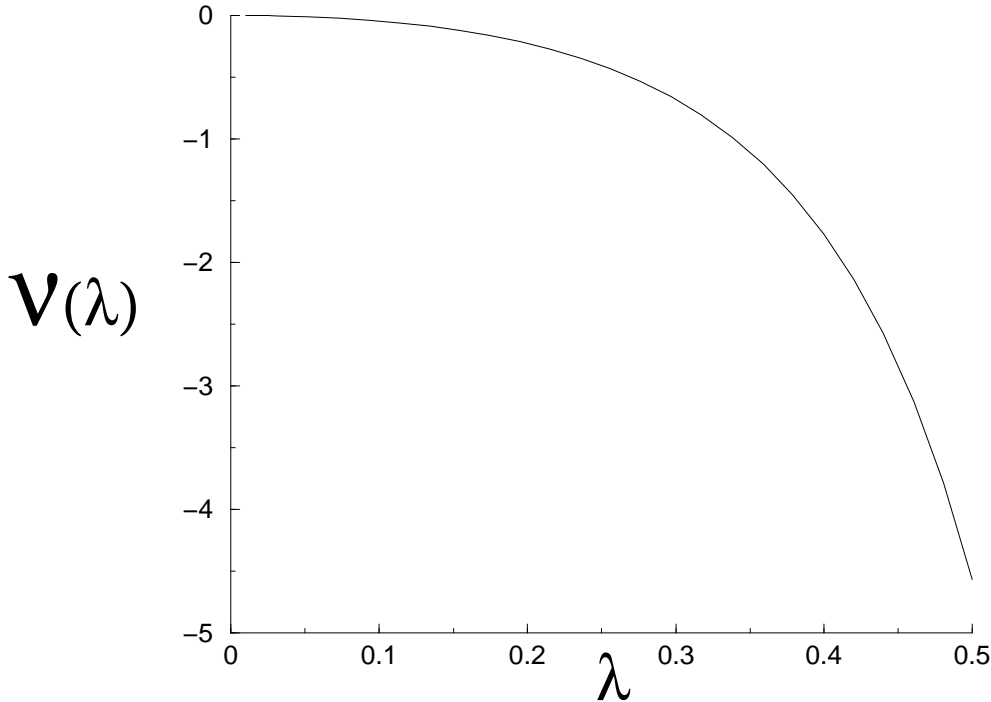}\\[3mm]
%\vspace*{6cm}
\label{fig3}
\caption[fig3]{The function ${\cal V}(\lambda)$.}
\end{figure}

The coefficients which accompany the
powers $\lambda^0$ and $\lambda^{D-1}$ will change
if we change the renormalization scale $\mu$ inside the logarithm. This
corresponds to the well known scaling of the effective potential
under changes of $\mu$, which 
adds a geometric term proportional to 
the coefficient $a_{D/2}(f=1,P)$ defined in (\ref{sdw}). It is easy to 
show that this coefficient is Weyl invariant and therefore independent 
of $\epsilon$. The invariant is made out of local terms (such as the 
square of the extrinsic curvature) which should also be included in the 
bare action. The renormalized couplings in front of these local terms 
are in fact supposed to be negligibly small (they have not been included 
in the equations of motion satisfied by the background), so the terms 
proportional to $\ln\mu$ might as well be dropped
from $V_0$. On the other hand, we can always add to $V_0$ terms which 
correspond to a finite renormalization of the cosmological constant in the 
bulk. These will give contributions proportional to $(1-\lambda^{D-1})$. 
Also, one can add terms which correspond to a finite renormalizaton of the 
tension on the branes. These will give independent
contributions proportional to $\lambda^{0}$ from
the positive tension brane and to $\lambda^{D-1}$ from the negative tension 
brane. Hence, the coefficients in front of these two powers of $\lambda$ 
cannot be determined from the calculation and can only be fixed by 
imposing suitable renormalization conditions which relate them to observables.

By using (\ref{conformal}) with the results (\ref{int1}) or (\ref{int2})
obtained in the preceding section, we can now evaluate $V^c_{\hbox{\footnotesize\it \hspace{-6pt} eff\,}}$. 
As mentioned at the end of the last subsection, 
the explicit calculation of the 
generalized
Seeley-De Witt coefficients is not strictly necessary. 
It is known that $a_{D/2}$ for odd $D$ 
has only contribution from the terms depending on the 
background quantities evaluated on the boundaries. Therefore, 
the shift in the effective potential due to the 
variation of $\epsilon$ must be given by the sum of 
a function of $z_+$ plus a function of $z_-$:
\begin{equation}
 {1\over {\cal A}_c}\int_0^1 d\epsilon\, 
   a_{D/2}(f_{\epsilon},P_{\epsilon})
  ={\cal F}_+(z_+) +{\cal F}_- (z_-).
\end{equation} 
Imposing that $V^c_{\hbox{\footnotesize\it \hspace{-6pt} eff\,}}$ has the correct scaling 
behaviour given in Eq. (\ref{scaling}), 
the functional form ${\cal F}_{\pm}$ is determined up to 
the two coefficients
in front of $\lambda^0$ and $\lambda^{D-1}$ mentioned above, 
which have the same $\lambda$ dependence as cosmological 
constants on the respective branes. 
Taking this into account, we have  
\begin{eqnarray}
 V^c_{\hbox{\footnotesize\it \hspace{-6pt} eff\,}}={1\over (4\pi)^{\eta}(\eta-1)! z_+^{D-1}}\Biggl[&& \alpha
           +\beta \lambda^{D-1}
         + \lambda^{D-1} {\cal V(\lambda)}
            \Biggr]. 
\label{efo}
\end{eqnarray}
The ``renormalized'' 
values $\alpha$ and $\beta$ can be obtained as before from the 
renormalization conditions (\ref{renc}).
An example of an effective potential after these conditions have been 
imposed is shown in Fig. 5.
Here 
$$
V_{\hbox{\footnotesize\it \hspace{-6pt} eff\,}}(\lambda)=a^{1-D}(z_+) 
  V^c_{\hbox{\footnotesize\it \hspace{-6pt} eff\,}}(\lambda),
$$
is the effective potential per unit physical volume on the positive tension
brane, which appears in Eq. (\ref{effect}).
(Note that, in fact, the conditions (\ref{renc}) do not depend on wether 
we are using the effective
potential per unit co-moving or per unit physical volume.)

\begin{figure}[t]
\centering
\hspace*{-4mm}
%\leavevmode
\epsfysize=10 cm \epsfbox{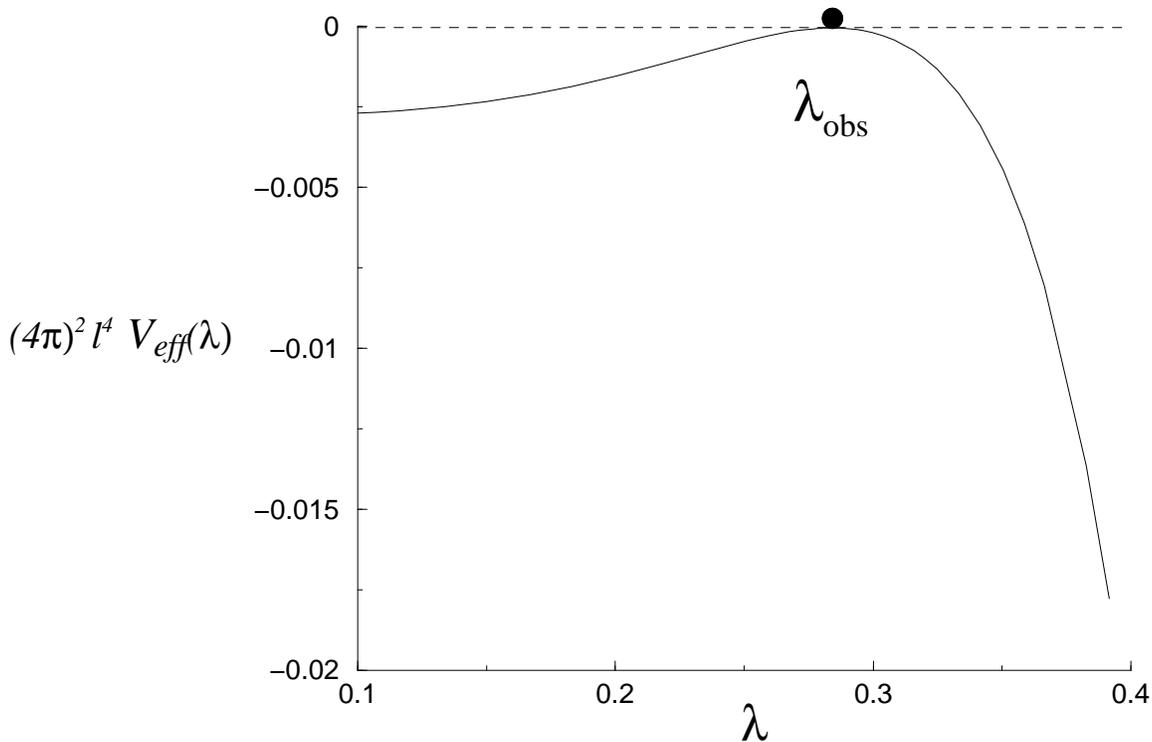}\\[3mm]
%\vspace*{6cm}
\label{fig4}
\caption[fig4]{Graviton contribution to the radion 
effective  potential.}
\end{figure}
 
Expanding ${\cal V}(\lambda)$ for small $\lambda$ and in $D=5$, we have
\begin{equation}
{\cal V}(\lambda)= {\cal I} \lambda^2 
   +O\left(\lambda^4\ln\lambda\right),
\end{equation}
where
\begin{equation}
{\cal I}=-{1\over 2}\int_0^{\infty} d\rho\,\rho^5
                {K_{1}(\rho)\over I_{1}(\rho)}
\approx -4.2 .
\end{equation} 
Hence, 
\begin{equation}
  V_{\hbox{\footnotesize\it \hspace{-6pt} eff\,}}(\lambda)
   ={\ell^{-4}\over (4\pi)^2}\left[\alpha 
                +\beta \lambda^4+{\cal I}\lambda^6+\cdots \right]. 
\label{finr}
\end{equation}
The second
equality in (\ref{renc}) determines $\beta\approx-(3/2) {\cal I} 
\lambda_{obs}^2$,
and then the first of the renormalization conditions gives the aditive 
constant $\alpha$. With this, the physical mass of the radion from the
point of view of the negative tension brane is given by
\begin{equation}
m^{2\ (-)}_\phi  
%=\lambda_{obs}^{-2}\ m^{2\ (+)}_\phi = 
% \lambda_{obs}^{-2}\ {d^2 V_{\hbox{\footnotesize\it \hspace{-6pt} eff\,}} \over
%  d\phi^2}
\approx \lambda_{obs}^{2}\left({{\cal I} \over\pi M^3 l^5}\right).
\label{massmin}
\end{equation}
For the case of the graviton considered here, this mass squared
is negative (this is shown here for small $\lambda_{obs}$, but it 
is in fact true for all $\lambda_{obs}$), and therefore the graviton cannot by 
itself stabilize the radion. 

As we have seen in Section II, in theories where additional 
fields can 
propagate in the bulk, some of them may give positive contributions to 
the mass squared. In fact, for a small $\lambda_{obs}$, the contribution
of conformally invariant fermions $m^{2\ (-)}_\phi \sim 
\lambda_{obs}\ (TeV)^2$, studied
in Section II, is much more important than that of gravitons, which scales 
as $\lambda_{obs}^2$.

\section{Conclusions and discussion}

We have shown that in brane-world scenarios with a warped extra dimension, 
it is in principle possible to stabilize the radion $\phi$ through the Casimir
force induced by bulk fields. In particular, conformally
invariant fields induce an effective potential of the form (\ref{confveff}) as
measured from the positive tension brane. From the point of view of the 
negative tension brane, this corresponds to an energy density per unit 
physical volume of the order
$$
V_{\hbox{\footnotesize\it \hspace{-6pt} eff\,}}^{-}\sim m_{pl}^4
\left[{A\lambda^4 \over (1-\lambda)^4}+\alpha+\beta\lambda^4\right],
$$
where $A$ is a calculable number (of order $10^{-3}$ per degree of freedom),
and $\lambda \sim \phi/(M^3 \ell)^{1/2}$ is the dimensionless radion. Here
$M$ is the higher-dimensional Planck mass, and $\ell$ is the AdS radius, 
which are both assumed to be of the same order, whereas $m_{pl}$ is the 
lower-dimensional Planck mass. In the absence of any fine-tuning,
the potential will have an extremum at $\lambda \sim 1$,
where the radion may be stabilized (at a mass of order $m_{pl}$).
However, this stabilization scenario without fine-tuning would not explain
the hierarchy between $m_{pl}$ and the $TeV$.

A hierarchy can be generated by adjusting $\beta$
according to (\ref{consts}), with $\lambda_{obs}\sim (TeV/m_{pl}) \sim 
10^{-16}$ (of course one must also adjust $\alpha$ in order to have vanishing
four-dimensional cosmological constant). But with these adjustement,
the mass of the radion would be very small, of order
\begin{equation} 
m^{2\ (-)}_{\phi} \sim\lambda_{obs}\ M^{-3} \ell^{-5} 
\sim \lambda_{obs} (TeV)^2.
\label{smallmass}
\end{equation}
Therefore, in order to make the model compatible with observations, 
an alternative mechanism must be invoked in order to  stabilize the radion,
giving it a mass of order $TeV$. 

Goldberger and Wise \cite{gw1,gw2},
for instance, introduced a field $v$ with suitable classical potential 
terms in the bulk and on the branes. In this model, the potential terms
on the branes are chosen so that the v.e.v. of the field in the positive
tension brane $v_+$ is different from the v.e.v. on the negative tension
brane $v_-$. Thus, there is a competition between 
the potential energy of the scalar field in the bulk and the 
gradient which is necessary to go from $v_+$ to $v_-$. The radion sits 
at the value where the sum of gradient and potential energies is 
minimized. This mechanism is perhaps somewhat {\em ad hoc}, but it has the 
virtue 
that a large hierarchy and an acceptable radion mass can be achieved without 
much 
fine tuning. It is reassuring that in this case the Casimir contributions,
given by (\ref{smallmass}), would be very small and would not spoil the model.

We have also calculated the graviton contribution to the radion effective
potential. Since gravitons are not conformally invariant, the calculation
is considerably more involved, and a suitable method has been developed for 
this purpose. The result is that gravitons contribute a negative term to
the radion mass squared, but this term is even smaller than (\ref{smallmass}),
by an extra power of $\lambda_{obs}$.

In an interesting recent paper \cite{FH}, Fabinger and Horava have considered
the Casimir force in a brane-world scenario similar to the one discussed 
in this paper, where the internal space is topologically $S^1/Z_2$. 
In their case, however, the gravitational field of 
the branes is ignored and the extra dimension is not warped. As a result, 
their effective potential is monotonic and stabilization does not occur 
(at least in the regime where the one loop calculation is reliable, just 
like in the original Kaluza-Klein compactification on a circle \cite{ac}). 
The question of gravitational backreaction of the
Casimir energy onto the background geometry is also discussed 
in \cite{FH}. Again, since the gravitational field of the branes is not 
considered, they do not find static solutions. This is in contrast with our
case, where static solutions can be found by suitable adjustment of the brane 
tensions. 

Finally, it should be pointed out that the treatment of backreaction
(here and in \cite{FH}) applies to conformally invariant fields but not to 
gravitons. Gravitons are similar to minimally coupled scalar 
fields, for which it is well known that the 
Casimir energy density diverges near the boundaries 
\cite{bida}. Therefore, a physical cut-off related 
to the brane effective width seems to be needed so that the energy density 
remains finite everywhere. Presumably, our conclusions will be 
unchanged provided that this cut-off length is small compared with the 
interbrane separation, but further investigation of this issue 
would be interesting.

\section{Acknowledgements}

We are grateful to Klaus Kirsten for useful discussions.
J.G. and O.P. acknowledge support from CICYT, under grant
AEN99-0766. O.P. is supported by a CIRIT, under grant
1998FI 00198. The stay of T.T. at IFAE was supported from 
Monbusho System to Send Researchers Overseas. 

\appendix

\section{Evaluation of the generalized zeta function}

In this Appendix, we evaluate the generalized zeta function (\ref{conver})
which appears in the expression of the effective potential (\ref{V0}).
We shall closely follow the method of Ref. \cite{lr2,lr1}.

The asymptotic behavior of the function $F$ in the integrand of 
(\ref{conver}) is given by 
\begin{equation}
 F(t)\to \left\{\begin{array}{l}
   \displaystyle 
   {i\over \pi\sqrt{\lambda} t} e^{-i(1-\lambda)t}\left(
       1+O(t^{-1})\right), \quad (i t \to \infty),\cr 
   \displaystyle 
   {-i\over \pi\sqrt{\lambda} t} e^{i(1-\lambda)t}\left(
       1+O(t^{-1})\right), \quad (i t \to -\infty). \end{array}
\right.
\end{equation}
Hence we rewite the zeta function as 
\begin{eqnarray}
 \hat\zeta(s) 
    &=&{s\over 2\pi i}\sum_{\sigma=\pm}\Biggl\{
        \int_{{\cal C}_{\sigma}} dt\, t^{-1-s}\ln\left[
          -\sigma\it\pi\sqrt{\lambda} t e^{\sigma i(1-\lambda)t} F(t)\right]
          -\int_{{\cal C}_{\sigma}} dt\, t^{-1-s}\ln\left[
          -\sigma\it\pi\sqrt{\lambda} t e^{\sigma i(1-\lambda)t}\right]
        \Biggr\}. 
\end{eqnarray}
Here, the contours ${\cal C}_{\pm}$ are the upper and lower halves of
the contour ${\cal C}$ in Fig. 3.

The contribution from the circle at infinity in the first term vanishes, 
so this part of the contour can be dropped. After that, the domain of 
convergence of the expression is extended to $\Re s>-1$. The second term can  
be explicitly evaluated for large $s$ as \begin{eqnarray}
 \int_{{\cal C}_{\pm}} &dt& t^{-1-s}\ln\left[
          \mp\it\pi\sqrt{\lambda} t e^{\pm(1-\lambda)t}\right]\cr
 &=& \mp\int_{\varepsilon}^{\infty} dt\, t^{-1-s}
          \ln\left[
          \mp\it\pi\sqrt{\lambda}t e^{\pm(1-\lambda)t}\right]\cr
 &=& \mp\left\{\left({\ln[\mp i\pi\sqrt{\lambda}]\over s}+{1\over s^2}\right)
         \varepsilon^{-s}+{1\over s}\varepsilon^{-s}\ln\varepsilon 
          \mp{i(1-\lambda)\over 1-s}\varepsilon^{1-s}\right\}.
\end{eqnarray}
Thus, the whole expression can be analytically continued  
to the strip given by $-1<\Re s<0$. 
Taking the $\varepsilon\to 0$ limit after this analytic 
continuation, we obtain 
\begin{eqnarray}
 \hat\zeta(s) 
    &=&{s\over \pi}\sin\left({\pi s\over 2}\right)
        \int_{0}^{\infty} d\rho\, \rho^{-1-s} \ln 
          \left[\pi\sqrt{\lambda}\rho e^{-(1-\lambda)\rho} 
           F(i\rho)\right]\cr
    & = & {s\over \pi}\sin\left({\pi s\over 2}\right)
        \int_{0}^{\infty} d\rho\, \rho^{-1-s} \ln 
        \left[2\sqrt{\lambda}\rho e^{-(1-\lambda)\rho}
         \left\{I_{\eta-1}(\rho)K_{\eta-1}(\lambda\rho)
             -I_{\eta-1}(\lambda\rho)K_{\eta-1}(\rho)\right\}\right]. 
\label{Hdef}
\end{eqnarray}
The asymptotic expansion of the modified Bessel functions is given by 
\begin{eqnarray}
  &&I_{\nu}(t)\sim {e^{t}\over \sqrt{2\pi t}} C_{\nu}(-t)
             +O(e^{-t}),\cr
  &&K_{\nu}(t)\sim \sqrt{\pi\over 2t} e^{-t} C_{\nu}(t),
\end{eqnarray}
with 
\begin{equation}
   C_{\nu}(t)= 
   \sum_{r=0}^{\infty}{\Gamma\left(\nu +r+{1\over 2}\right)\over r!\,
            \Gamma\left(\nu -r+{1\over 2}\right)}  (2t)^{-r}. 
\end{equation}
We define coefficients $\beta_r$ by 
\begin{equation}
   \ln\left[ C_{\eta-1}(\rho)\right]=
   \sum_{r=1}^{\infty} {\beta_r \over \rho^r}.
\end{equation}
Now, we rewrite the expression for $\hat\zeta(s)$ as 
\begin{eqnarray}
\hat\zeta(s)= {s\over \pi}\sin\left({\pi s\over 2}\right)\Biggl\{
   &&     \lambda^s\int_{0}^{\infty} dt\, t^{-1-s} 
          \ln\left[\sqrt{2t/\pi}\, e^{t}K_{\eta-1}(t)\right]\cr
   &&   +  \int_{0}^{\infty} d\rho\, \rho^{-1-s} 
          \ln\left[\sqrt{2\pi\rho} e^{-\rho}I_{\eta-1}(\rho)\right]\cr
 &&     +  \int_{0}^{\infty} d\rho\, \rho^{-1-s} 
          \ln\left[1-{I_{\eta-1}(\lambda\rho)\over 
                 K_{\eta-1}(\lambda\rho)}{K_{\eta-1}(\rho)\over 
                 I_{\eta-1}(\rho)}\right]\Biggr\}. 
\label{zetaTama}
\end{eqnarray}
The first integral in the square brackets can be 
evaluated in the following way. 

Defining the function 
\begin{equation}
R(t)=\sum_{r=1}^{D-2} {\beta_r\over t^r}
         +{\beta_{D-1}\over t^{D-1}} e^{-\left|{1\over t}\right|}, 
\end{equation}
we can rewrite that integral as  
\begin{eqnarray}
\int_{0}^{\infty} dt\, t^{-1-s} 
          \ln\left[\sqrt{2t/\pi}\, e^{t}K_{\eta-1}(t)\right]
 &=&\int_{0}^{t_0} dt\, t^{-1-s} 
          \ln\left[\sqrt{2t/\pi}\, e^{t}K_{\eta-1}(t)\right]
\cr &&+\int_{t_0}^{\infty} dt\, t^{-1-s} 
          \left(\ln\left[\sqrt{2t/\pi}\, e^{t}K_{\eta-1}(t)\right]
             -R(t)\right)
\cr &&+\sum_{r=1}^{D-2}{\beta_r\over s+r}t_0^{-s-r} 
 +\beta_{D-1}\int_{t_0}^{\infty} dt\, t^{-s-D} e^{-{\left|1\over t\right|}}.  
\label{IK}
\end{eqnarray}
Except for poles, this expression is now analytic in the strip 
$-(D-2)<\Re s<0$.
The integral in the last term is given by
\begin{eqnarray}
\int_{t_0}^{\infty} dt\, t^{-s-D} e^{-{1\over t}}
  &= & 
\int_{0}^{1\over t_0} dx\, x^{s+(D-1)-1} e^{-x}\cr
 &=&{1\over s+(D-1)}\left\{
     t_0^{-s-(D-1)} e^{-{1\over t_0}}
     +\int_{0}^{1\over t_0} dx\, x^{s+(D-1)} e^{-x}
      \right\}. 
\end{eqnarray}
After substitution of this expression in (\ref{IK}), we can 
perform analytic continuation to $-(D-2)> \Re s > -D$. 
Then, taking the $t_0\to 0$ limit, the right hand side of Eq. (\ref{IK})
becomes 
\begin{equation}
\int_{0}^{\infty} dt\, t^{-1-s} 
          \left(\ln\left[\sqrt{2t/ \pi}\, e^{t}K_{\eta-1}(t)\right]
             - R(t)\right)
     +{\beta_{D-1}\Gamma(s+D) \over s+(D-1)}. 
\end{equation}
Using $\Gamma(1+x)=1-\gamma x+O(x^2)$, we find 
\begin{equation}
\int_{0}^{\infty} dt\, t^{-1-s} 
          \ln\left[\sqrt{2t/\pi}\, e^{t}K_{\eta-1}(t)\right]
   ={\beta_{D-1}\over s+(D-1)} +{\cal I}_K +O\left(s+(D-1)\right),
\end{equation}
where
\begin{equation}
 {\cal I}_K=-\gamma \beta_{D-1}+\int_{0}^{\infty} dt\, t^{D-2} 
          \left(\ln\left[\sqrt{2t/\pi}\, e^{t}K_{\eta-1}(t)\right]
             - R(t)\right). 
\end{equation}

The second term in the square brackets in (\ref{zetaTama}) can 
be evaluated in a similar way as 
\begin{equation}
\int_{0}^{\infty} d\rho\, \rho^{-1-s} 
          \ln\left[\sqrt{2\pi\rho} e^{-\rho}I_{\eta-1}(\rho)\right]
   ={\beta_{D-1}\over s+(D-1)} +{\cal I}_I +O\left(s+(D-1)\right), 
\end{equation}
where
\begin{equation}
 {\cal I}_I=-\gamma \beta_{D-1}+\int_{0}^{\infty} d\rho\, \rho^{D-2} 
          \left(\ln\left[\sqrt{2\pi\rho}\; e^{-\rho}I_{\eta-1}(\rho)\right]
             - R(-\rho)\right).  
\end{equation}

Substituting the above results into (\ref{zetaTama}), we obtain 
\begin{equation}
 \hat\zeta(1-D)=-(-1)^{\eta}\eta \beta_{D-1}
        (1+\lambda^{1-D})=\left\{\begin{array}{l}
   \displaystyle
   {1\over 16}(1+\lambda^{-2}), \quad (D=3),\cr
   \displaystyle
   {27 \over 64} (1+\lambda^{-4}), \quad (D=5), \end{array}
\right.
\end{equation}
and 
\begin{equation}
 \hat\zeta'(1-D)=-(-1)^{\eta}\eta\Biggl[ {\cal I}_I+
      \lambda^{1-D}{\cal I}_K
+\beta_{D-1} \left({1+\lambda^{1-D} \over 1-D}
+ \lambda^{1-D}\ln\lambda\right)
         + {\cal V}(\lambda) \Biggr],
\end{equation}
where ${\cal V}(\lambda)$ is given in (\ref{calvo}).
Then, substituting these results into (\ref{V0}), 
we obtain (\ref{result}).

\end{document}